\documentstyle[aps,multicol,preprint]{revtex}
\tightenlines 
\begin{document}
\title{
Dirac's hole theory versus quantum field theory }
\author
{F.A.B. Coutinho$^a$, D. Kiang$^b$, Y. Nogami$^c$ and Lauro Tomio$^d$}
\address
{$^a$Faculdade de Medicina, Universidade de S\~ao Paulo,  
Av. Dr. Arnaldo 455, \\ 01246-903 S\~ao Paulo, Brazil
 (E-mail: coutinho@dns2.fm.usp.br) \\
$^b$Department of Physics, Dalhousie University, Halifax, 
Nova Scotia, Canada B3H 3J5\\
 (E-mail: dkiang@fizz.phys.dal.ca)\\
$^c$Department of Physics and Astronomy, McMaster University,  
Hamilton, Ontario, \\ Canada L8S 4M1
 (E-mail: nogami@mcmaster.ca)\\
$^d$Instituto de F\'{\i}sica Te\'orica, Universidade Estadual
Paulista, Rua Pamplona, 145, \\ 01405-900, S\~ao Paulo, Brazil
 (E-mail: tomio@ift.unesp.br)}
\date{\today}
\maketitle
\begin{abstract}
Dirac's hole theory and quantum field theory are usually
considered equivalent to each other. For models of a certain 
type, however, the equivalence may not hold as we discuss 
in this Letter. This problem is closely related to the validity 
of the Pauli principle in intermediate states of perturbation 
theory. 
\newline\newline
\end{abstract}

\vskip 1cm

\section{introduction}
Dirac's hole theory (HT) \cite{1} and quantum field theory 
(QFT) are usually considered equivalent to each other. 
This equivalence, however, does not necessarily hold for 
models of a certain type as noted recently \cite{2}. The 
purpose of this Letter is to elaborate on this possible 
inequivalence. We start with the Dirac equation for a 
particle in a given potential in single particle quantum 
mechanics. We examine how the energy eigenvalues of the 
Dirac equation vary when an external perturbation is 
applied. Then we consider HT and QFT which are based on 
the same Dirac equation. We focus on the vacuum state and 
its energy shift caused by an external perturbation. 

The possible inequivalence between HT and QFT is closely 
related to the validity of Feynman's prescription to 
disregard the Pauli principle (PP) in intermediate states 
of perturbation theory \cite{3}. This prescription is 
based on Feynman's observation that effects of all virtual 
processes that violate PP cancel out (at least formally). 
In HT, however, the PP-violating terms do not necessarily 
cancel as explicitly illustrated in \cite{4}. In such a 
case, the result of perturbation calculation differs 
depending on whether or not PP is enforced in intermediate 
states. Then the question arises: Should we enforce PP or
not?

Cavalcanti \cite{5} found the exact solution of the HT 
model of \cite{4} in its special case with the particle
mass $m=0$. He pointed out that the exact solution is 
consistent with the perturbation calculation in which PP 
is disregarded throughout but not with the one in which 
PP is enforced. This was puzzling in view of the usual
belief that PP operates in intermediate states. 
Cavalcanti's comment was responded by \cite{2} but the 
above puzzling aspect and the difference between HT and 
QFT were not fully addressed in it. In this Letter we 
emphasize that for models of a certain type, including 
that of \cite{4}, HT and QFT are not necessarily 
equivalent and we explore various aspects of the 
difference. In QFT, the cancellation of the PP-violating 
terms in perturbation theory is complete. QFT is free 
from the puzzling aspect that was found in \cite{4,5}.  

In Section II we specify the type of the model that we 
consider. We discuss some difficulties of the HT version 
of the model. In Section III we examine the QFT version 
of the model and clarify the difference between HT and QFT. 
In Section IV we raise a further question regarding HT.   

\section{The hole theory}
We start with single-particle relativistic quantum 
mechanics with the Hamiltonian 
\begin{equation}
H=H_0 + V,
\label{1}
\end{equation}
where $H_0$ is the Dirac Hamiltonian for a particle in a 
binding potential and $V$ is an external perturbation. 
Unperturbed and perturbed systems, respectively, are such 
that
\begin{equation}
H_0 \phi_n = \epsilon_n \phi_n,\;\;\;\;
H\psi_n= \eta_n \psi_n.
\label{2}
\end{equation}
Suffix $n$ ($=\pm1,\pm2,\ldots$) specifies eigenstates. 
The energy eigenvalues are labeled such that 
$0<\epsilon_1 < \epsilon_2 < \ldots\,$, and 
$0>\epsilon_{-1} > \epsilon_{-2} > \ldots\,$., and 
similarly for $\eta_n$. We assume that the energy levels 
are all discrete and non-degenerate and that there is a 
one-to-one correspondence between unperturbed and 
perturbed eigenstates. In other words, $\eta_n \to 
\epsilon_n$ when $V \to 0$. Let us also assume that 
$\eta_n$ and $\epsilon_n$ have the same sign. It is not 
difficult to relax these restrictions.

We now turn to HT and consider the vacuum in which all 
negative-energy states are occupied. The unperturbed and 
perturbed energies of the vacuum are respectively given by 
$E_0 = \sum_{j} \epsilon _{-j}$ and $E = \sum_{j} \eta _{-j}$. 
We are interested in the energy shift due to the perturbation,
\begin{equation}
\Delta E = \sum_{j}(\eta _{-j} -\epsilon _{-j}).
\label{3}
\end{equation}
Suppose that we start with known $\epsilon_n$ and attempt
to reach $\eta_n$ by perturbation theory. We follow the 
standard prescription of perturbation theory including 
{\em all} intermediate states. We treat the system as a 
one-body system and we do not exclude the intermediate 
states which are already occupied by other particles. The 
first order energy shift is given by
\begin{equation}
\Delta E^{(1)} = \sum_j V_{-j,-j} 
= \int V({\bf r})\rho_{\rm vac} ({\bf r}) d{\bf r},
\label{4}
\end{equation} 
where $V_{-j,-j}=\langle \phi_{-j}|V|\phi_{-j}\rangle $ and 
$\rho_{\rm vac} ({\bf r})=\sum_j |\phi_{-j}({\bf r})|^2$ is 
the particle density of the unperturbed vacuum. This 
$\Delta E^{(1)}$ as such generally diverges because 
$\rho_{\rm vac}({\bf r})$ is actually infinite. No matter 
how weak it is, $V$ may cause an infinite energy shift. This 
difficulty can be avoided by assuming that the density 
in the perturbed vacuum itself is not an observable quantity 
and that only the difference between the density of the
perturbed vacuum and its unperturbed counterpart is 
observable. Then the first order energy shift disappears.

We are more interested in the second order energy shift, 
\begin{equation}
\Delta E^{(2)} = \sum_j \Delta {\epsilon^{(2)}_{-j}}
=\sum_j \sum_i \frac{|V_{i,-j}|^2}
{\epsilon_{-j} -\epsilon_i} + X,
\label{5}
\end{equation}
\begin{equation}
X = \sum_j \sum_{k\neq j} \frac{|V_{-k,-j}|^2}
{\epsilon_{-j} -\epsilon_{-k}},
\label{6}
\end{equation}
where $V_{i,-j}=\langle \phi_i |V|\phi_{-j} \rangle$. Term 
$X$ is due to the transitions between negative energy states 
like the one from $-j$ to $-k$. In $\Delta E^{(2)}$,
PP has not been considered. If we enforce PP, the transitions 
between negative energy states are not allowed and we obtain
\begin{equation}
\Delta E^{(2)}_{\rm PP} = \sum_j \sum_i \frac{|V_{i,-j}|^2}
{\epsilon_{-j} -\epsilon_i},
\label{7}
\end{equation} 
where suffix PP means ``with PP enforced". The $X$ of Eq.
(\ref{6}) is an infinite alternating series. This formally 
vanishes because the numerator is symmetric with respect to 
$j \to k$ while the denominator is antisymmetric. As was 
illustrated in \cite{4}, however, $X$ may not vanish 
depending on how the calculation is done.

In \cite{4}, the one-dimensional bag model with 
$H_0 =\alpha p+\beta [m + S(x)]$ and perturbation 
$V(x)=\lambda x$ was considered. Here $m$ is the mass of
the particle, $p=-i\hbar d/dx$, $\alpha = \sigma_y$ and 
$\beta =\sigma_z$ are $2\times 2$ Dirac matrices. The
$S(x)$ is a Lorentz-scalar potential such that $S(x)=0$
for $|x|<a$ and $S(x)=\infty$ for $|x|>a$. In this model, 
we obtain $V_{-j,-j}=0$ and $\Delta E^{(1)}=0$. The 
divergence difficulty of $\Delta E^{(1)}$ that we pointed 
out below (\ref{4}) does not arise. For the second order 
energy shift the calculation was done in two ways, methods 
I and II. In method I, PP is enforced whenever it is 
applicable. In method II, PP is disregarded throughout. 
Method I leads to $\Delta E^{(2)}_{\rm PP}$ and method II 
to $\Delta E^{(2)}$. The explicit calculations of \cite{4} 
led to $\Delta E^{(2)}_{\rm PP}<0$ and $\Delta E^{(2)}=0$.

Cavalcanti \cite{5} showed that, if $m=0$, the perturbed 
Dirac equation $H\psi_n= \eta_n \psi_n$ for the model of 
\cite{4} can be solved analytically. This enabled him to 
obtain the exact energy shift of the model. He showed that 
$\eta =\epsilon$. The eigenvalue $\eta$ is independent of 
the perturbation and so is the exact energy of the HT 
vacuum, that is, $\Delta E =0$. This is consistent with 
$\Delta E^{(2)}=0$ of method II but not with 
$\Delta E^{(2)}_{\rm PP}<0$ of method I. In other words, 
if one enforces PP in intermediate states, one obtains a 
wrong energy shift in HT. This is puzzling. According to 
the spin-statistics theorem (which was proved in QFT), the 
wave function of a fermion system has to be antisymmetric 
with respect to interchanges of the particles. This implies 
PP. The perturbative interaction $V$ is totally symmetric 
with respect to the interchange of particles. It acts in 
the same way on all particles. Then the intermediate state 
that is connected to the antisymmetric initial state through 
$V$ has to be antisymmetric. Hence PP is expected to hold 
in the intermediate states. This is also related to the 
unitarity of the $S$ matrix \cite{6,7}. Recall also that 
Dirac introduced HT such that the vacuum state is stable. 
This stability relies on PP.

Cavalcanti solved the Dirac equation by rewriting it 
as $dw_{\pm}(x)/dx \mp i(\lambda x -\eta)w_{\mp}(x) =0$,
where $w_{\pm}(x) = u(x)\pm iv(x)$, $u(x)$ and $v(x)$ 
being the upper and lower components of $\psi(x)$,
respectively. His solution is of the form of
$w_{\pm}(x)=C_{\pm}e^{\pm i[(\lambda/2)x^2-\eta x]}$
where $C_{\pm}$ are constants. His solution can be 
generalized to the case with an arbitrary potential $V(x)$ 
(but still with $m=0$) by replacing his $w_{\pm}(x)$ with
\begin{equation}
w_{\pm}(x)=C_{\pm}e^{\pm i[f(x)-\eta x]},\;\;\; 
\frac{df(x)}{dx}=V(x).
\label{8}
\end{equation}
The boundary condition for the wave function is 
$w_+ (\pm a)=\mp i w_- (\pm a)$, which leads to
\begin{equation}
\eta-\epsilon=\frac{1}{2a}\int_{-a}^a V(x)dx,
\label{9}
\end{equation}
where $\epsilon=(2n+1)(\pi/4a)$ with $n=0, \pm 1, \pm 2, 
\, \dots$, is the eigenvalue when $V(x)=0$. It is
remarkable that this exact energy-shift of the 
single-particle state is first order. The second order 
and all higher order effects are zero. 

If $V(x)$ is an odd function of $x$, then $\eta =\epsilon$.
This leads to $\Delta E =0$, that is, the exact energy of 
the HT vacuum is independent of $V(x)$. If $V(x)$ contains 
an even function part, all eigenvalues are shifted exactly 
by the same amount. The vacuum of HT contains an infinite 
number of negative-energy particles. If the energy of every 
particle is shifted by the same amount, the total energy of 
the vacuum obtains an infinite energy shift. This 
illustrates what we pointed out below Eq. (\ref{4}).

\section{Quantum field theory}
In QFT the Hamiltonian for the unperturbed system can 
be expressed as 
\begin{equation}
{\cal H}_0 = \sum _i \epsilon_i a^{\dagger}_i a_i
+ \sum _j \bar\epsilon_j b^{\dagger}_j b_j,
\;\;\; \bar\epsilon_{j} = -\epsilon_{-j} >0. 
\label{10}
\end{equation}
The notation is hopefully self-explanatory. The $a_i$ 
($a^{\dagger}_i$) is the creation (annihilation) operator 
for the particle in the unperturbed state $i$.  The 
$a^{\dagger}_i$ creates a particle with energy $\epsilon_i$ 
with wave function $\phi _i (x)$. These operators satisfy 
the usual anticommutation relations. The $b_j$ 
($b^{\dagger}_j$) is for the antiparticle. No 
negative-energy particles appear in QFT. Let us emphasize 
that the Hamiltonian is defined in terms of normal products 
of the creation and annihilation operators. The unperturbed 
vacuum contains no particles nor antiparticles. It is  
the eigenstate of ${\cal H}_0$ with zero eigenvalue. 

The interaction Hamiltonian is of the form of
\begin{equation}
{\cal V} 
= \sum_{i,i'} V_{i,i'}a^{\dagger}_i a_{i'}
+ \sum_{i,j} V_{i,-j}a^{\dagger}_i b^{\dagger}_j
+ \cdots\,,
\label{11}
\end{equation}
where, for example, $a^{\dagger}_i b^{\dagger}_j$ creates 
a pair of particle and antiparticle. The other terms that
are not shown above are of the form of $ab$, $a^{\dagger}b$,
$b^{\dagger}a $ and $b^{\dagger} b$. The number of the 
particles is conserved with the understanding that an 
antiparticle has particle number -1. Because ${\cal V}$ 
consists of normal products, its expectation value in the 
unperturbed vacuum is zero. Hence we obtain
\begin{equation}
\Delta {\cal E}^{(1)} = 0.
\label{12}
\end{equation}
We use notation ${\cal E}$ for the energy of the system 
in QFT. Equation (\ref{12}) is in a sharp contrast to 
Eq. (\ref{4}). QFT is free from the difficulty of the 
infinite first-order energy shift of HT that we pointed 
out below (\ref{4}). If we did not use the normal products 
for ${\cal V}$, we would obtain (\ref{4}) for 
$\Delta {\cal E}^{(1)}$.

Next let us examine the energy shift of the second order.
In QFT the vacuum does not contain any particles nor
antiparticles. The transitions between negative
energy states of HT has no place in QFT. We obtain
\begin{equation}
\Delta {\cal E}^{(2)} = -\sum_j \sum_i \frac{|V_{i,-j}|^2}
{\epsilon_i +\bar \epsilon_j},
\label{13}
\end{equation}
which agrees with $\Delta E^{(2)}_{\rm PP}$ of HT. Recall 
that $\Delta E^{(2)}_{\rm PP}$ and $\Delta E^{(2)}$ may 
or may not agree with each other in HT.

If the exact solutions of the perturbed Dirac equation
are known, they can be used to define the creation and 
annihilation operators $c$, $c^\dagger$, $d$ and 
$d^\dagger$ for the perturbed system. The Hamiltonian 
for the perturbed system then becomes
\begin{equation}
{\cal H} = \sum _i \eta_{i} c^{\dagger}_{i} c_{i}
+ \sum _j \bar\eta_{j} d^{\dagger}_{j} d_{j},
\;\;\; \bar\eta_{j} = -\eta_{-j} >0 .
\label{14}
\end{equation}
The perturbed vacuum is the eigenstate of ${\cal H}$ 
with zero eigenvalue. Perturbed vacuum contains 
no particles nor antiparticles which are defined in
terms of $\{ c,c^{\dagger},d,d^{\dagger}\}$.

In the absence of perturbation, the vacuum energy is zero.
In the presence of perturbation, if we use ${\cal H}$ given 
above, the vacuum energy is again zero. This may give the 
false impression that the perturbation causes no energy 
shift. Recall that the Hamiltonian is defined as a normal 
product. The normal product depends on how the creation and 
annihilation operators are defined. This dependence gives 
rise to the energy shift. This can be seen as follows. We 
have two sets of basis functions, $\{ \phi_n \}$ and 
$\{ \psi_n \}$, each of which forms a complete orthonormal 
system. The $\{ a,a^{\dagger},b,b^{\dagger}\}$ and 
$\{ c,c^{\dagger},d,d^{\dagger}\}$ are related by the
transformation,
\begin{equation}
c_i =\sum_{i'}\langle \psi_i|\phi_{i'}\rangle a_{i'}
+ \sum_j \langle \psi_i|\phi_{-j}\rangle b^{\dagger}_j \, ,
\label{15}
\end{equation}
\begin{equation}
d^{\dagger}_j =\sum_{i}\langle \psi_{-j}|\phi_{i}
\rangle a_{i}
+ \sum_{j'} \langle\psi_{-j}|\phi_{-j'}\rangle
b^{\dagger}_{j'} \, ,
\label{16}
\end{equation}
and their hermitian adjoints. If we rewrite 
${\cal H}_0 +{\cal V}$ in terms of 
$\{ c,c^{\dagger},d,d^{\dagger}\}$, we expect to obtain
\begin{equation}
{\cal H}_0 + {\cal V} = {\cal H} + \Delta {\cal E}.
\label{17}
\end{equation}
The $\Delta {\cal E}$ is the expectation value of
${\cal H}_0 + {\cal V}$ in the perturbed vacuum. 
The $\Delta {\cal E}$ is also the negative of the 
expectation value of ${\cal H}$ in the unperturbed vacuum.
If ${\cal V}\to 0$, then ${\cal H}\to {\cal H}_0$
and $\Delta {\cal E}\to 0$. Therefore $\Delta {\cal E}$ 
is the energy shift of the vacuum due to the perturbation 
${\cal V}$.

Equation (\ref{17}) has various interesting implications. 
Let us first work out $\Delta {\cal E}$ explicitly. It 
is somewhat simpler to start with ${\cal H}$ and rewrite 
it in terms of $\{ a,a^{\dagger},b,b^{\dagger}\}$. The 
combination $c^{\dagger}_{i} c_{i}$ goes like
\begin{eqnarray}
c^{\dagger}_{i} && c_{i} = \left( \sum_{i'}
\langle \psi_i|\phi_{i'}\rangle ^* {a^\dagger}_{i'}
+ \sum_j \langle \psi_i|\phi_{-j}\rangle ^* b_j \right)
\left(\sum_{i'}\langle \psi_i|\phi_{i'}\rangle
a_{i'}
+\sum_j \langle \psi_i|\phi_{-j}\rangle b^{\dagger}_j \right)
\nonumber \\ && 
= ({\rm normal\;\;products\;\;of\;}\, a, a^{\dagger},\, 
\dots) + |\langle\psi_i |\phi_{-j}\rangle|^2 ,
\label{18}
\end{eqnarray}
and similarly for $d^{\dagger}_{i} d_{i}$. If we
substitute the above into ${\cal H}$ and collect the
terms of the form of, e.g., $a^\dagger a$, we obtain
\begin{eqnarray}
\sum_{i',i"} && \left( \sum_{i}\eta_i 
\langle \psi_i|\phi_{i'}\rangle ^* 
\langle \psi_i|\phi_{i"}\rangle 
-\sum_{j}\bar\eta_j \langle \psi_{-j}|\phi_{i'}\rangle 
\langle \psi_{-j}|\phi_{i"}\rangle ^* \right)
{a^\dagger}_{i'} a_{i"} \nonumber \\
&&=\sum_{i,i'}
\langle \phi_{i}|(H_0 + V)|\phi_{i'}\rangle 
{a^\dagger}_{i} a_{i'} 
= \sum_i \epsilon_i {a^\dagger}_{i} a_{i}
+ \sum_{i,i'}V_{i,i'}{a^\dagger}_{i} a_{i'}.
\label{19}
\end{eqnarray}
In this way the normal products of 
$a,\, a^{\dagger},\, \dots$ altogether become 
${\cal H}_0 + {\cal V}$ and we obtain
\begin{equation}
\Delta {\cal E} = -\sum_i \sum_j 
\left(\eta_i |\langle\psi_i |\phi_{-j}\rangle|^2
+ \bar\eta_j |\langle\psi_{-j}|\phi_i \rangle|^2\right) .
\label{20}
\end{equation}
Equation (\ref{20}) we believe is new. 
The $\Delta {\cal E}$ can vanish only if ${\cal V}=0$. 
Otherwise it is negative. Let us emphasize that this 
sign of $\Delta {\cal E}$ is not related to the sign 
of the single-particle energy-shift $\eta - \epsilon$. 
Even if $\eta -\epsilon$ is positive for all levels, 
$\Delta {\cal E}$ is negative. This does not
necessarily mean though that perturbations ${\cal V}$
and $-{\cal V}$ lead to the same value of 
$\Delta {\cal E}$.

The $\Delta {\cal E}$ is the exact energy shift. Let us 
examine its leading term in the perturbation expansion. 
It is obvious that $\Delta {\cal E}^{(1)}=0$. For the
matrix elements involved, we obtain in first order 
\begin{equation}
\langle \psi_i |\phi_{-j}\rangle =\frac{V_{i,-j}}
{\epsilon_{i}-\epsilon_{-j}}, \;\;\;\;
\langle \psi_{-j} |\phi_{i}\rangle = \frac{V_{-j,i}}
{\epsilon_{-j}-\epsilon_{i}}.
\label{21}
\end{equation}
As far as the leading term of $\Delta {\cal E}$ is 
concerned, we can put $\eta_i =\epsilon_i$ and 
$\bar \eta_j =\bar\epsilon_j =-\epsilon_{-j}$. Then 
follows Eq. (\ref{13}). One can also work out higher
order energy shifts $\Delta {\cal E}^{(3)}$, etc.,
successively.

The problem regarding PP in intermediate states does 
not arise in the QFT calculation for the vacuum 
energy. There is nothing that blocks the particle and 
antiparticle pair creation in the second-order 
intermediate states. Because the transitions between 
negative energy states do not appear in QFT, the 
perturbation calculation is free from the ambiguity 
of the kind that was pointed out in \cite{4}. The two 
methods of calculation I and II lead to exactly the 
same results in QFT. All the PP-violating effects in 
method II cancel. One can disregard PP throughout as 
Feynman advocated.

In HT which is based on the one-dimensional bag model 
with $m=0$, there are no second-order and higher order 
effects in the energy shift. In contrast to this, 
the exact energy shift of the QFT vacuum is given by 
$\Delta {\cal E}$ of Eq. (\ref{20}). Its first-order 
part is absent whereas its second-order part is negative 
definite. The exact energy shift can be explicitly 
calculated by using the known exact solutions of the 
Dirac equation of the model.

Let us point out a feature of Eq. (\ref{17}) which 
is interesting in relation to Feynman's prescription. 
Consider a system that consists of a number of particles 
and antiparticles together with their vacuum background. 
The energy shift of this system is simply given by
\begin{equation}
\sum_i (\eta_i -\epsilon_i) +
\sum_j (\bar\eta_j -\bar\epsilon_j)+\Delta {\cal E},
\label{22}
\end{equation}
where the summations are for the particles and 
antiparticles. The $\Delta {\cal E}$ is the same as that
of Eq. (\ref{20}). In calculating $\eta_i -\epsilon_i$,
for example, one does not have to think about the
vacuum background. On the other hand $\Delta {\cal E}$ 
is the energy shift of the vacuum in the absence of the
additional particles and antiparticles. This can be 
illustrated by using the QFT version of the 
one-dimensional bag model of \cite{4}.

There is another interesting feature of Eq. (\ref{17}).
The energy spectra of the system in the sense of the
differences between energy levels are simply determined
by the single particle energies $\eta$'s and $\bar\eta$'s.
The vacuum energy $\Delta {\cal E}$ is common to all
energy levels of the system.

\section{Discussion}
We started with a problem of single-particle quantum 
mechanics with the Dirac equation for a particle in a
given potential. We then considered the HT and QFT
versions of the problem and examined how the vacuum 
energy shifts when an external perturbation is applied. 
We discussed a situation such that HT and QFT are not 
equivalent. In case of such discrepancy we should 
choose QFT rather than HT.

If HT and QFT are not necessarily equivalent, some of the
calculations done in the framework of HT may have to be
reexamined. As one of such possible problems, let us 
mention the fractional fermion number. In HT, an external 
perturbation causes a change in the density distribution
in the vacuum in which all negative energy levels are 
filled. This may result in fractionalization of the 
fermion number of the vacuum; See, e.g., \cite{8,9,10,11}. 
In QFT the unperturbed vacuum is empty. When an external 
perturbation is applied to it, a pair of particle and 
antiparticle can be created. This process of pair creation, 
however, does not change the fermion number of the vacuum. 
It is not clear whether or not this can be compatible with 
the fermion number fractionalization that occurs in HT.

\section*{Acknowledgements}
We would like to thank Dr. A. Ni\'{e}gawa for enlightening
discussions. This work was supported by Funda\c c\~{a}o de 
Amparo \`{a} Pesquisa do Estado de S\~{a}o Paulo (FAPESP), 
Conselho Nacional de Desenvolvimento Cient\'\i fico e 
Tecnol\'{o}gico (CNPq) and the Natural Sciences and 
Engineering Research Council of Canada. 
\nopagebreak

\end{document}